\documentclass[aps,prl,prb,twocolumn,showpacs,superscriptaddress,groupedaddress]{revtex4-2}
\usepackage{graphicx}
\usepackage{physics}
\usepackage{amssymb}
\usepackage{amsmath}
\usepackage{latexsym}
\usepackage{ulem}
\usepackage{color}
\usepackage{dcolumn}
\usepackage{subfigure}
\usepackage[T1]{fontenc}
\usepackage[utf8]{inputenc}
\usepackage[english,french]{babel}
\usepackage[colorlinks=true,linkcolor=blue,citecolor=blue]{hyperref} %
\usepackage{bm}        % for math
\usepackage{amssymb}   % for math

\usepackage[capitalize]{cleveref}

\begin{document}

\graphicspath{{figures/}}

\definecolor{airforceblue}{rgb}{0.36, 0.54, 0.66}

\title{Flat band superconductivity in a system with a tunable quantum metric: \\ the stub lattice}

\author{M. Thumin}%
 \email{maxime.thumin@neel.cnrs.fr}
 \author{G. Bouzerar}%
 \email{georges.bouzerar@neel.cnrs.fr}
\affiliation{Université Grenoble Alpes, CNRS, Institut NEEL, F-38042 Grenoble, France}%
\date{\today}

\begin{abstract}
Over the past years, one witnesses a growing interest in flat band (FB) physics which has become a playground for exotic phenomena. In this study, we address the FB superconductivity in one-dimensional stub chain. In contrast to the sawtooth chain or the creutz ladder, for a given strength of the attractive electron-electron interaction, the stub chain allows the tuning of the real space spreading of the FB eigenstates (quantum metric or QM). We study in detail the interplay between the interaction strength and the mean value of the QM $\langle g \rangle$ on the pairings and on the superfluid weight $D_s$. Our calculations reveal several interesting and intriguing features. For instance, in the weak coupling regime, $D_s$ with respect to $\langle g \rangle$ exhibits two different types of behaviour. Despite the fact that the pairings differs drastically, $D_s$ scales linearly with the QM only when its $\langle g \rangle$ is large enough (small gap limit). On the other hand, when the QM is of small amplitude an unusual power law is found, more precisely $D_s \propto \langle g \rangle ^\nu$ where $\nu \rightarrow 2$ in the limit of large single particle gap.  
In addition to the numerical calculations, we have provided several analytical results which shed light on the physics in both the weak and strong coupling regime. Finally, we have addressed the impact of the thermal fluctuations on the superfluid weight.

\end{abstract} 

\maketitle

%%%%%%%%%%%%%%%%%%%%%%%%%%%%%%%%%%%
%%%%%%%%%%%%%%%%%%%%%%%%%%%%%%%%%%%
%%%%%%%%%% DEBUT %%%%%%%%%%%%%%%%%%
%%%%%%%%%%%%%%%%%%%%%%%%%%%%%%%%%%%
%%%%%%%%%%%%%%%%%%%%%%%%%%%%%%%%%%%

\section{Introduction}

For the last two decades, the interest in flat bands (FB) material has been growing a lot, placing this emerging family of compounds at the heart
of the physics of strongly correlated systems \cite{Synthese_FB,Frac_Hall_1,Frac_Hall_2,Frac_Hall_3,Lieb_Lattice_Origine,Tasaki_magnetism}.
Due to destructive quantum interference, the eigenstates can be localized \cite{Localization}, leading to a constant energy band over the whole Brillouin zone (BZ). The kinetic energy being quenched, the interaction energy becomes the unique relevant energy scale, and exotic phases of quantum matter can emerge in such materials. On top of those quantum phases stands the superconductivity which has been intensively studied lately. Experimentally, superconducting phases which are very likely of FB origin have been reported in graphene based material such as the twisted bilayer graphene (TBG) \cite{Cao2018_1,Cao2018_2, Efetov} as well as in graphite \cite{HOPG_2000, HOPG_2012}, while theoretical studies have covered a wide range of low dimensional systems. Despite the Mermin-Wagner theorem \cite{Mermin_Wagner, Hohenberg}, two-dimensional systems such as the TBG \cite{Torma_TBG}, the Lieb lattice \cite{Peotta_Lieb, Batrouni_CuO2} or the dice lattice \cite{Dice_chinois} often considered as systems in which a superconducting phase transition of topological nature can occur without spontaneous continuous symmetry breaking. It corresponds to the Berezinsky-Kosterlitz-Thouless (BKT) transition \cite{Berezinsky_1972, Kosterlitz_1972, Kosterlitz_1973}. 
More recently, one-dimensional systems are getting under the spotlights \cite{Batrouni_Creutz, Batrouni_sawtooth, Batrouni_Designer_Flat_Bands, Peotta_SU(2)}. Indeed, one and quasi-one-dimensional systems are good candidates to facilitate the understanding of the underlying physics and may as well be relevant for the superconductivity in anisotropic systems \cite{Quasi_1D_Cs2Cr3As3, Quasi_1D_Nature, Quasi_1D_NaMn6Bi5}. In FB superconductors, the superfluid weight has two kind of contribution: a conventional intraband component (vanishing in the strictly FB limit), and an interband term of geometric nature. In the weak coupling regime, the geometric contribution varies linearly with the quantum metric (QM) tensor as defined in Ref. \cite{Origine_Quantum_Metric} and with the interaction strength \cite{Peotta_Nature}.

\begin{figure}[h!]
\centering
\includegraphics[scale=0.35]{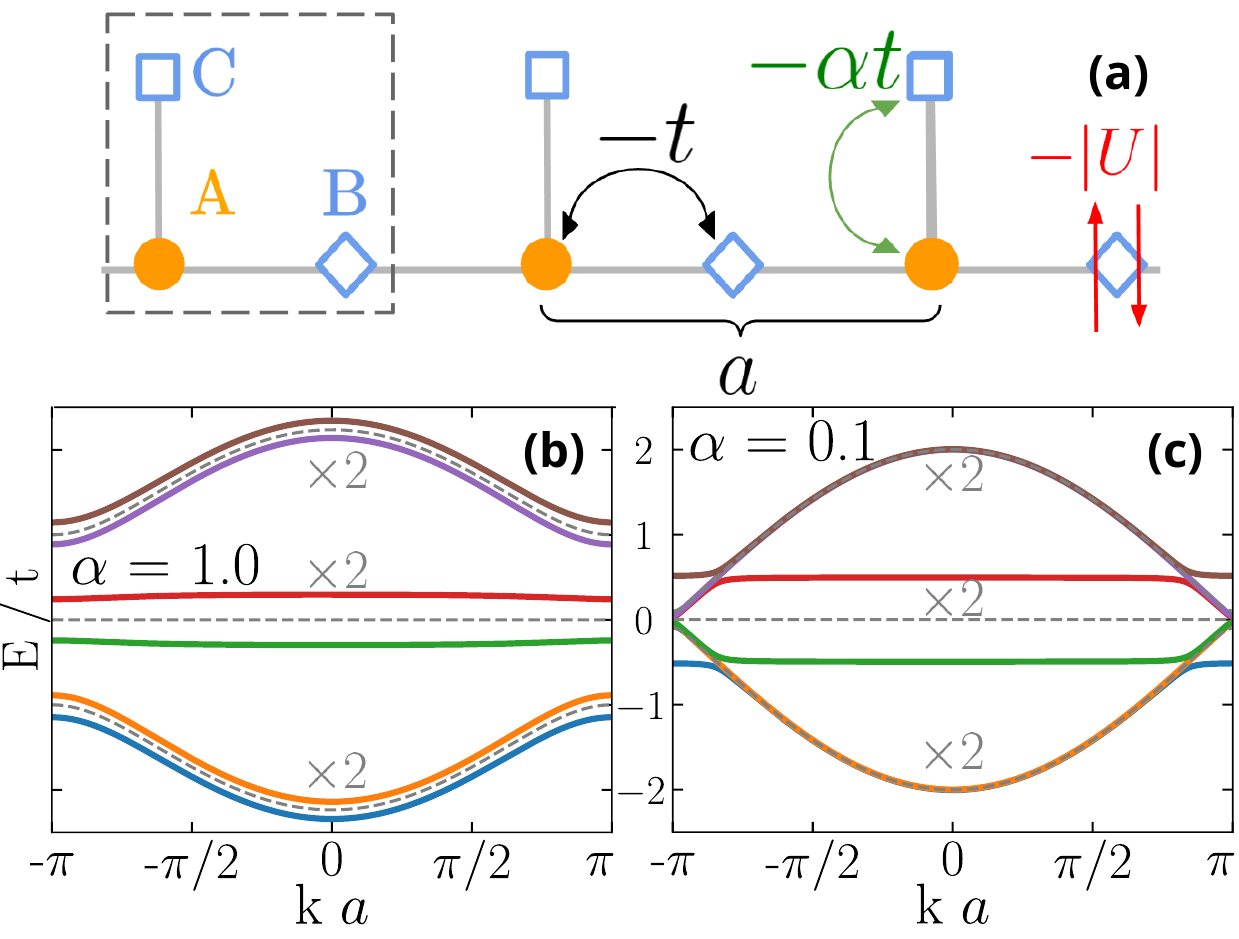} 
\caption{\textbf{(a)} Schematic representation of the attractive Hubbard Hamiltonian (Eq.$\,$\eqref{H_exact} in the main text) for electrons in the stub lattice. Quasi-particle dispersions for $|U|/t=0$ (dashed lines) and $|U|/t=1$ (continuous lines) with respectively $\alpha=0.1$ \textbf{(b)} and $\alpha$ = 1 \textbf{(c)}. The symbol '$\cross 2$' means that the eigenvalues are twofold degenerate.}
\label{Fig. 1}
\end{figure}

The purpose of the present work is to consider a FB system where the QM is tunable. The lattice chosen to pursue this study is a bipartite chain with 3 atoms per unit cell so-called the stub lattice, as illustrated in Fig$\,$\ref{Fig. 1}\textbf{a}. Unlike the sawtooth chain or the Creutz ladder, the stub chain is bipartite and hosts a FB for any value of the out-of-chain hopping, $\alpha t$ in Fig.\ref{Fig. 1}\textbf{a}, which provides the freedom to tune the QM. 
Despite its absence of natural realization, one should mention that the artificial stub lattice can be experimentally engineered, for instance, within the optical lattice framework \cite{stub_photonics}, or even by the realization of micro-pillar optical cavities \cite{stub_optical_cavities}.

\section{Model and methods}

Electrons in the stub lattice, in the presence of attractive electron-electron interaction, are described by the Hubbard model,
\begin{equation}
\begin{split}
    \hat{H} = \sum_{\langle i\lambda,j\eta \rangle, \sigma} t^{\lambda\eta}_{ij} \; \hat{c}_{i\lambda, \sigma}^{\dagger} \hat{c}_{j \eta, \sigma} - \mu \hat{N} - |U| \sum_{i\lambda} \hat{n}_{i\lambda\uparrow}\hat{n}_{i\lambda\downarrow}
    \label{H_exact}
\end{split}
,
\end{equation}
where the operator $\hat{c}_{i \lambda \sigma}^{\dagger}$ creates an electron of spin $\sigma$ at site $\textbf{r}_{i\lambda}$, $i$ being the cell index and $\lambda=$A,B and C. The sums run over the lattice, $\langle i\lambda,j \eta\rangle$ refers to nearest-neighbor pairs for which the hopping integral $t^{\lambda\eta}_{ij}$ is $t$ for (AB) pairs and $\alpha t$ for (AC) pairs. $\hat{N}=\sum_{i\lambda,\sigma} \hat{n}_{i\lambda,\sigma}$ is the particle number operator, $\mu$ is the chemical potential, and finally $|U|$ is the strength of the on-site attractive electron-electron interaction. In what follows, the lattice spacing $a$ will be set to $1$.

To address the FB superconductivity in the stub chain, we propose to handle the electron-electron interaction term within the mean-field Bogoliubov-De-Gennes (BdG) approach. Before we proceed, it is crucial to justify the relevance and accuracy of BdG as compared to methods such as exact diagonalization (ED), density matrix renormalisation group (DMRG), Quantum-Monte-Carlo (QMC) and dynamical mean field theory (DMFT). 
In the case of the Lieb lattice, a good agreement was found between BdG and ED calculations of the superfluid weight $D_s$ \cite{Peotta_Lieb}. Similarly, in the case of FB superconductivity in CuO$_2$ layers, BdG has fairly reproduced the pair structure factor as obtained in the QMC simulations \cite{Batrouni_CuO2}. Moreover, in one-dimensional systems, such as the sawtooth chain, the creutz ladder and other quasi-one dimensional FB systems, the calculation of $D_s$ by BdG and DMRG has revealed an impressive quantitative agreement \cite{Batrouni_sawtooth, Batrouni_Designer_Flat_Bands}. Furthermore, We should mention that the BCS wavefunction is the exact ground-state for any bipartite lattice hosting FB while the FB is gapped and $|U|$ is smaller than the gap \cite{Peotta_Nature, Peotta_Lieb}. In addition, we quote as well the fact that the mean field unrestricted Hartree-Fock theory has been shown to be very accurate to describe the magnetic phases of strongly correlated electrons in two-dimensional decorated lattices which exhibit quasi-FB in the vicinity of the Fermi energy \cite{Ralko_Bouzerar}.
Thus, one can confidently and safely consider that the BdG approach is a suitable and reliable tool to address quantitatively the FB superconductivity in the stub lattice.  

Before discussing our results, we recall briefly the BdG theory. The U-term is decoupled as follows,
$\hat{n}_{i\lambda,\uparrow}\hat{n}_{i\lambda,\downarrow} \longrightarrow 
 \langle\hat{n}_{i\lambda,\downarrow}\rangle _{th} \hat{n}_{i\lambda,\uparrow} + \langle\hat{n}_{i\lambda,\uparrow}\rangle _{th} \hat{n}_{i\lambda,\downarrow}
 - \frac{\Delta_{i\lambda}}{|U|} \hat{c}^{\dagger}_{i\lambda,\uparrow}\hat{c}^{\dagger}_{i\lambda,\downarrow} 
- \frac{\Delta^*_{i\lambda}}{|U|}\hat{c}_{i\lambda,\downarrow}\hat{c}_{i\lambda,\uparrow} - C_{i\lambda}$ where $\Delta_{i\lambda} = -|U|\langle\hat{c}_{i\lambda,\downarrow}\hat{c}_{i\lambda,\uparrow}\rangle _{th}$ are the pairing order parameters, and $C_{i\lambda} = \langle\hat{n}_{i\lambda,\uparrow}\rangle _{th} \langle\hat{n}_{i\lambda,\downarrow}\rangle _{th} + \langle\hat{c}_{i\lambda,\downarrow}\hat{c}_{i\lambda,\uparrow}\rangle _{th} \langle\hat{c}^{\dagger}_{\lambda,\uparrow}\hat{c}^{\dagger}_{i\lambda,\downarrow}\rangle _{th}$. 
For a fixed temperature and a given density of electrons, the pairings and the occupations are calculated self-consistently. 
Notice that translation invariance implies that the thermal average $\langle\ldots\rangle _{th}$ of a local operator is cell-independent. Thus, we drop the cell index. We consider as well a paramagnetic ground-state,  $\langle\hat{n}_{\lambda,\uparrow}\rangle_{th}=\langle\hat{n}_{\lambda,\downarrow}\rangle_{th}=\langle\hat{n}_{\lambda}\rangle_{th}/2$.
Eq.$\,$\eqref{H_exact} becomes, 
\begin{equation}
    \hat{H}_{BdG} = \sum_k 
    \begin{bmatrix}
        \hat{c}_{k \uparrow}^{\dagger} & \hat{c}_{-k \downarrow} \\ 
        \end{bmatrix} 
       \begin{bmatrix} 
         h^{\uparrow}(k)& \hat{\Delta} \\ 
         \hat{\Delta}^{\dagger}& -h^{\downarrow}(-k) \\
   \end{bmatrix}
          \begin{bmatrix} 
         \hat{c}_{k \uparrow} \\ 
         \hat{c}_{-k \downarrow}^{\dagger} \\
   \end{bmatrix}
   ,
\end{equation}

where $\hat{c}_{k\sigma}^{\dagger} = \Big(\hat{c}_{k A,\sigma}^{\dagger}, \hat{c}_{kB,\sigma}^{\dagger}, \hat{c}_{kC,\sigma}^{\dagger}\Big)$, $c_{k\lambda,\sigma}^{\dagger} $ is the Fourier transform (FT) of $c_{i\lambda,\sigma}^{\dagger}$.  $\hat{h}^\sigma (k)=\hat{h}_0(k)-\mu-\hat{V}_{\sigma}$ where $\hat{h}^\sigma_0$ is the FT of the tight-binding term in Eq.$\,$\eqref{H_exact}, $\hat{V}_\sigma=\frac{|U|}{2}\,\text{diag}(\langle\hat{n}_{A}\rangle_{th},\langle\hat{n}_{B}\rangle_{th},\langle \hat{n}_{C} \rangle_{th})$ and $\hat{\Delta}=\text{diag}(\Delta_A, \Delta_B, \Delta_C)$.

\section{Results and discussions}

\subsection{Quasi-particles dispersions}

In the present study we focus our attention on the half-filled case for which $\mu=-|U|/2$ and $n_\lambda = 1$ as it is predicted by the uniform density theorem in bipartite lattices \cite{Th_Lieb_uniform}.
In Fig.$\,$\ref{Fig. 1}\textbf{b,c} are plotted the quasi-particle (QP) dispersions for $|U|/t=0$ and $|U|/t=1$, with respectively $\alpha=0.1$ and $\alpha=1$. First, for $U=0$, a gap $\delta_0$ of amplitude $|\alpha|t$ opens up in the one particle spectrum between the FB and the dispersive bands at $k=\pi$ (bands are degenerate). When $U$ is switched on, the degeneracy of each band is lifted. For small values of $\alpha$ ($\alpha=0.1)$, we observe pronounced differences between the vicinity of $k=0$ and $k=\pi$. The splitting of the high energy bands is significant in vicinity of $k=\pi$, whilst in the rest of the Brillouin zone (BZ) it is negligible. On the other hand, the former FB remain flat except near the BZ boundary where it behaves as a massive Dirac excitation, with a small QP gap $\Delta_{QP}$ of the order of $0.025\,t$ ($\alpha=0.1$) for $|U|/t=1$. Notice that the splitting between the quasi-FBs is of the order of $|U|$ at the zone center. 
 In contrast, for larger values of $\alpha$, the high energy bands splitting is almost k-independent and the former FBs are quasi-flat in the whole BZ.
Notice, that the splitting of the quasi-FB at $k=0$ is smaller than for $\alpha=0.1$, i.e. $0.29~|U|$.

\subsection{Pairings and quasi-particles gap}

\begin{figure}[h!]
    \centering
\includegraphics[scale=0.4]{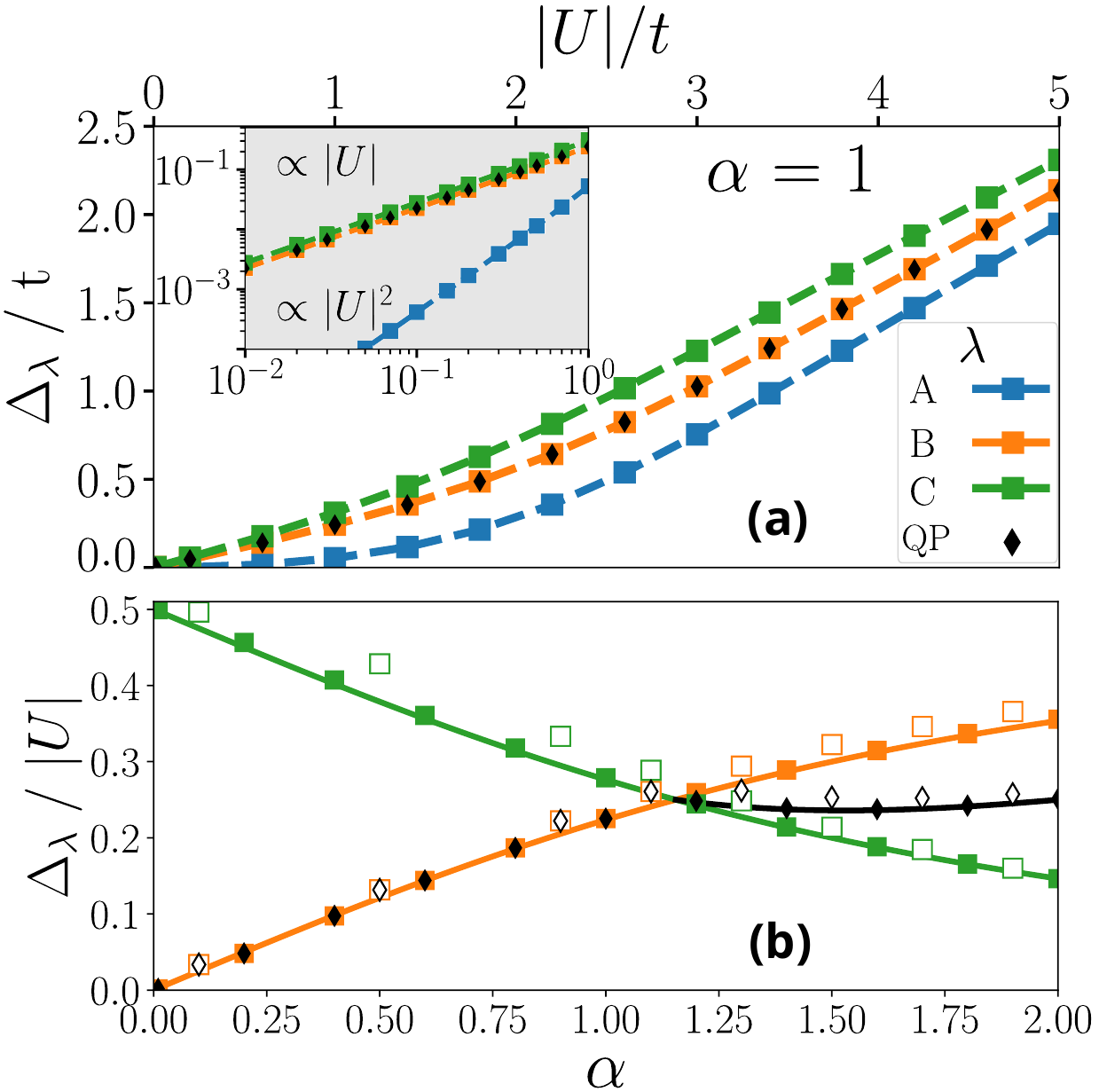}    
     \caption{\textbf{(a)} Pairings ($\lambda = A,B,C$) and quasi-particle gap ($\lambda=QP$) as a function of $|U|$ for $\alpha=1$. The weak interaction region is magnified in the inset. \textbf{(b)} $\frac{\Delta_\lambda}{|U|}$ as a function of $\alpha$ where the symbols represent the numerical data. The filled (resp. empty) symbols correspond to $|U|=0.1\,t$ (resp. $|U|=t$) and the continuous lines are the analytical calculations.}
 \label{Fig. 2}
\end{figure}

Fig.$\,$\ref{Fig. 2}\textbf{(a)} depicts the pairings and $\Delta_{QP}$ as a function of $|U|$ for $\alpha=1$. Note that the pairings are taken real, since they all have the same phase which can be removed by global gauge transformation. For small $|U|$, both $\Delta_B$ and $\Delta_C$ scale linearly with $|U|$ and $\Delta_A \propto|U|^2$. Such a behavior is consistent with what has been reported in recent studies \cite{Peotta_Nature, Batrouni_Creutz} and as it has been pointed out in former studies \cite{Gap_linear_1990, Gap_linear_1994, BCS_FB}. It will be discussed in more details in what follows.
This scaling contrasts with the conventional BCS theory which predicts $\Delta_{BCS} \propto t\,e^{-1/|U|\rho(E_F)}$ for the half-filled one-dimensional chain \cite{BCS}. As anticipated, in the strong coupling regime ($|U|\gg t$), the pairing increases linearly with $|U|$.
 In addition, $\Delta_\lambda$ is found orbital-independent and $\Delta_\lambda \simeq \frac{|U|}{2}$ as expected for the half-filled system when the charge density is uniform. In Fig.$\,$\ref{Fig. 2}\textbf{(b)} is plotted $\Delta_\lambda/|U|$ as a function of $\alpha$ for $|U|\leqslant t$. The numerical data are obtained for both $|U|=0.1~t$ and $|U|= t$. For small values of $\alpha$, we find $\Delta_B \propto \ |U| \alpha$ and $\Delta_C \approx \frac{|U|}{2}$ which can be understood by considering the expression of the FB compact localized eigenstate (CLS) that reads $\ket{\text{CLS}_i} = \frac{1}{\sqrt{2+\alpha^2}}(\ket{C_i} + \ket{C_{i+1}} - \alpha \ket{B_i} )$. In this regime, the weight is roughly constant on C-sites and varies linearly with $\alpha$ on B-sites. Thus, as $\alpha$ increases, $\Delta_C$ decays, and simultaneously $\Delta_B$ rises until they finally cross at $\alpha_c \approx 1.2 \pm 0.1$, where the CLS weight is comparable on both B and C sites.
In addition, Fig.$\,$\ref{Fig. 2}\textbf{b} reveals two distinct regimes for the QP gap. More specifically, for $\alpha \leqslant \alpha_c$, the gap is located at $k=\pi$ and $\Delta_{QP}=\Delta_B$. On the other hand, for $\alpha \geqslant \alpha_c$, it moves to $k=0$, it is weakly $\alpha$-dependent and lies between $\Delta_B$ and $\Delta_C$. Within a first order perturbation calculation with respect to $|U|$(see Appendix A), one gets the following set of analytical 
 expressions,
\begin{equation}
\begin{split}
    \Delta_B = \frac{|U|\alpha}{2\sqrt{4+\alpha^2}}, \\
    \Delta_C+\Delta_B = \frac{|U|}{2}.
\end{split}
\end{equation}
Thus, the sum $\Delta_C+\Delta_B$ is $\alpha$-independent. We find as well for the QP gap,
\begin{equation}
    \Delta_{QP} = \left\{ 
\begin{array}{l l}
 \qquad \Delta_B \qquad \, \text{for}  \quad \alpha \leqslant \alpha_c \\
 \frac{\alpha^2 \Delta_B+4\Delta_C}{\alpha^2+4} \quad  \text{for} \quad \alpha \geqslant \alpha_c, \\ 
 \end{array} 
 \right.   
\end{equation}
where $\alpha_c=\frac{2}{\sqrt{3}}= 1.155$.
As it can be seen, Fig.$\,$\ref{Fig. 2}\textbf{b} nicely illustrates the excellent (resp. good) quantitative agreement between the numerical calculations for $|U|=0.1~t$ (resp. $|U|= t$) and the analytical calculations.

\subsection{Superfluid weight and quantum metric}

The SC phase is characterized by the superfluid weight \cite{Khon_Ds, Shastry_Sutherland_Ds, Scalapino_Ds} defined as 
\begin{equation}
D_s=\frac{1}{ N_c}\frac{\partial^2 \Omega (q)}{\partial q^2} \Big|_{q=0} ,
\end{equation}
where $N_c$ is the number of unit cell of the lattice, $\Omega (q)$ is the grand-potential and $q$ mimics the effect of a vector potential, introduced by a standard Peierls substitution $t^{\lambda\eta}_{ij} \longrightarrow t^{\lambda\eta}_{ij} e^{iq(x_{i\lambda}-x_{j\eta})}$. 

\begin{figure}[h!]
    \centering
    \includegraphics[scale=0.4]{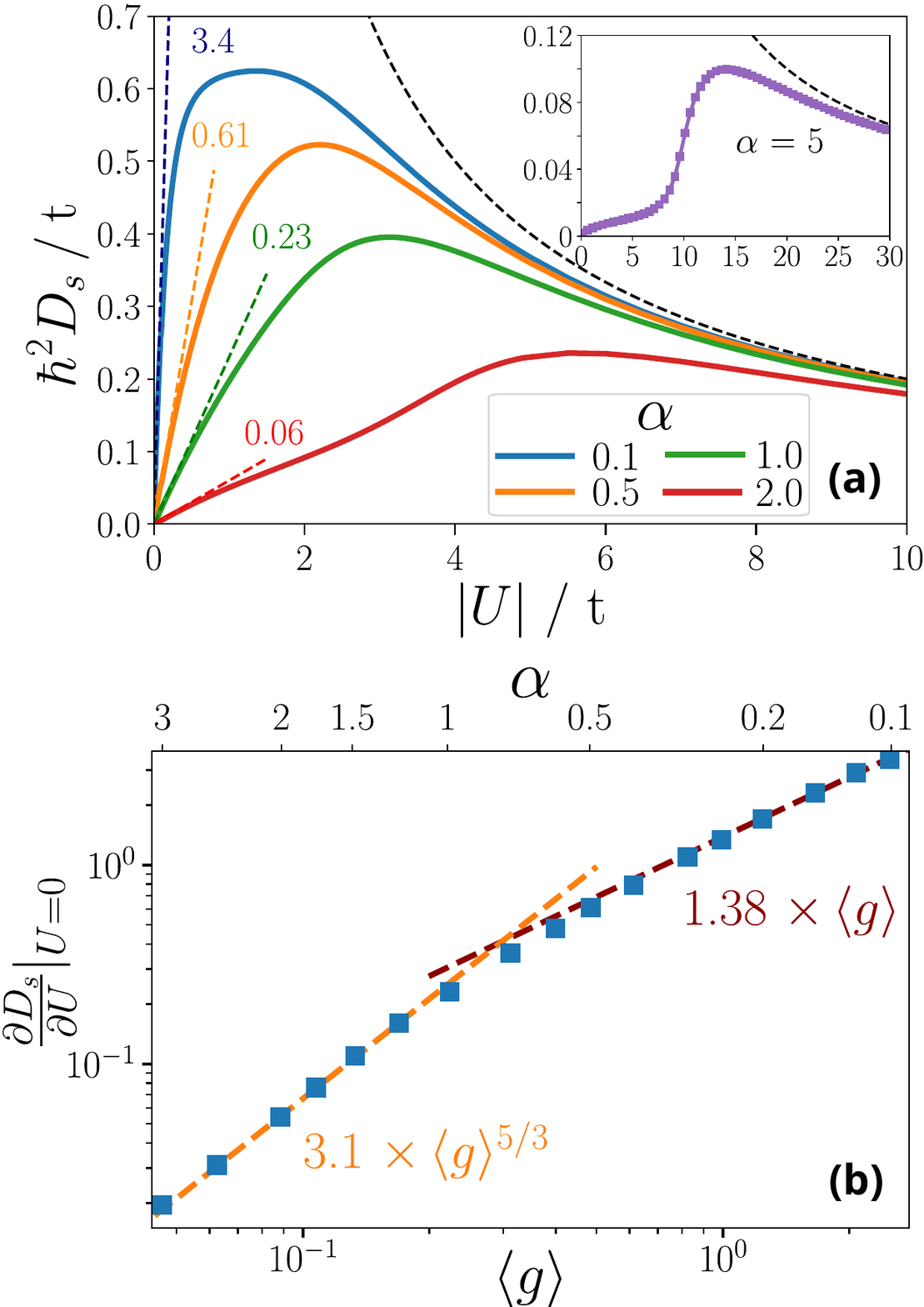} 
    \caption{\textbf{(a)} Superfluid weight $D_s$ at $T=0$ as a function of $|U|$ for $\alpha= 0.1, 0.5, 1$ and $2$. The inset represents $D_s$ for a large value of $\alpha$. The black dashed line is the analytical expression in the limit $|U| \gg t$ (see text). \textbf{(b)} $\frac{\partial D_s}{\partial U} |_{U=0}$ as a function of $\langle g \rangle$, the mean value of the quantum metric (square symbols). The corresponding values of $\alpha$ are depicted on the upper $x$-axis. The dashed lines are data fits discussed in the main text.}
    \label{Fig. 3}
\end{figure}

Fig.$\,$\ref{Fig. 3}\textbf{a} depicts $D_s$ as a function of $|U|$ for different values of $\alpha$. We first consider the low U region where one observes that $D_s \propto |U|$ as it has been established for isolated FB \cite{Peotta_Nature}. Starting from $\alpha=1$ and as we reduce it, the slope increases very rapidly. We find $\frac{\partial D_s}{\partial |U|}=0.23$, $0.61$ and $3.4$ for respectively $\alpha=1$, $0.5$ and $0.1$. Simultaneously, the region where $D_s \propto |U|$ shrinks significantly as $\alpha$ decreases.
Additionally, as $\alpha$ increases beyond $\alpha=1$, the slope is now drastically suppressed, e.g. for $\alpha=2$ it drops to $0.06$. Fig.$\,$\ref{Fig. 3}\textbf{b} illustrates the connection between $D_s$ and the mean value of the quantum metric (QM) of the FB eigenstates defined as $\langle g \rangle = \frac{1}{2\pi} \int_{-\pi}^{\pi} dk g(k)$, where we recall the definition of the QM \cite{Origine_Quantum_Metric},
\begin{equation}
g(k) = 
\braket{\partial_k \psi^{FB}_{k}}{\partial_k \psi^{FB}_{k}} - |\braket{\psi^{FB}_{k}}{\partial_k \psi^{FB}_{k}} |^2
\end{equation}
$\ket{\psi^{FB}_k}$ is the FB eigenstate of the non-interacting Hamiltonian. For the stub lattice, one finds $g(k) = \frac{\sin{(\frac{k}{2})}}{\alpha^2 + 4 \, \text{cos}^2{(\frac{k}{2})}}$ which leads to $\langle g \rangle = \frac{1}{2|\alpha|\sqrt{2+\alpha^2}}$. 

For isolated half-filled FBs, within BdG approach it has been shown analytically that $D_s=2|U|n_{\phi}\langle g \rangle$, $n_{\phi}^{-1}$ being the number of orbitals on which the FB wave function is finite \cite{Peotta_SU(2)}. The validity and accuracy of this result has been for instance confirmed numerically by DMRG for the Creutz ladder \cite{Tovmasyan_Peotta, Batrouni_Creutz}.

In the limit of vanishing U, we find two distinct types of behaviour. For $\alpha \leqslant\alpha_c$ the SF weight scales linearly with the QM and a fit of the plotted data gives a ratio $R=\frac{D_s}{\langle g \rangle} \approx 1.38$. Notice that according to Ref.\cite{Peotta_Nature} and with $n_{\phi}^{-1}=2$, one would find $R=1$. From our analytical calculations, available in the Appendix C, in the regime $\alpha \ll \alpha_c$ a ratio $R=3/2$ has been found. On closer inspection, this is intriguing. Recall that to obtain $\frac{D_s}{U} \propto \langle g \rangle$ it requires (i) a uniform pairing on the sites where the CLS weight is finite and (ii) a large gap ($\delta_0 \gg |U|$) between the dispersive bands and the FB. While condition (ii) is fulfilled, the first one is not. Indeed, in the limit $\alpha \ll 1$, the ratio $\Delta_B / \Delta_C$ is of the order of $\alpha$ (see Fig.$\,$\ref{Fig. 2}\textbf{b}) which means as well that the pairing occurs essentially on C-sites. Hence, for a finite $|U|$, one would expect instead a vanishing superfluid weight as $\alpha$ goes to zero. 

It raises the crucial question of how to resolve these contradictions.
For this, notice that the square root of the mean value of the QM provides a measure of the mean spread of the FB eigenstates \cite{Marzari_spread_PRB, Marzari_sprea_revew, Giant_boost}. More precisely the QM can be re-expressed, 
\begin{equation}
    g(k) = \bra{\psi_k^{FB}}(\hat{x}^2 - \langle \hat{x} \rangle _k ^2) \ket{\psi_k^{FB}},     
\end{equation}
where $\hat{x}$ is the position operator and $\langle \hat{x} \rangle _k = \bra{\psi_k^{FB}}\hat{x}\ket{\psi_k^{FB}}$. This leads, for $\alpha \ll 1$, to a mean spread of the FB eigenstates $\Bar{L} = \sqrt{\langle g \rangle} = \frac{1}{2\sqrt{\alpha}}$. 
From a dimensional point of view, the SF weight is the ratio of a typical energy scale $\delta E$ of the quasi-FB (QFB) divided by the the square of a typical momentum $q_{typ}$. 
For small U, the bandwidth $\delta W$ of the QFB is the relevant energy scale. As shown in the Appendix A, in the limit of vanishing $\alpha$, this bandwidth is $\alpha$-independent and $\delta W = \frac{|U|}{2}$. On the other hand, the natural choice for $q_{typ}$ is $\frac{1}{\Bar{L}}$, since there is no Fermi wave vector for flat bands. Thus, the SF weight should scale as,
\begin{equation}
D_s \sim \delta W \times \Bar{L}^2,
\end{equation}
which can be as well rewritten $D_s \sim |U| \langle g \rangle$.  

On the other hand, for $\alpha \geqslant \alpha_c$, the data show that $\frac{D_s}{|U|}$ is inconsistent with a linear dependence of the QM. A fit of the numerical data suggest an unusual scaling, $D_s = 3.1|U|\langle g \rangle^{\nu}$, where $\nu \approx 1.7$. However, it is found that the power law is sensitive
to the region chosen for the fit. Moreover, the convergence becomes more difficult as $\alpha$ becomes too large.
Based on our numerical data, for $\alpha \gg 1$, $\nu$ seems to converge to 2. Using arguments similar to those discussed above, one can also explain this change of behavior. For large $\alpha$, the bandwidth of the QFB is now $\alpha$-dependent and falls out rapidly as $\alpha$ increases. More precisely, it is found that $\delta W = \frac{2|U|}{\alpha^2}$ as shown in the Appendix A, and from the QM expression one has $\Bar{L}^2 = \frac{1}{2\alpha^2}$, yielding $D_s \sim \frac{|U|}{\alpha^4} \propto |U| \langle g \rangle ^2$. This scaling is confirmed by the detailed analytical calculations available in the Appendix C.

We now propose to discuss the intermediate and strong coupling regime. For any $\alpha$, the shape of $D_s$ as a function of $|U|$ is similar, namely, after a linear increase with respect to $|U|$, the SF weight reaches a maximum $D^{max}_s$ and then decays monotonously as $|U|$ gets larger. The location of the maximum $U_{max}$ strongly depends on $\alpha$ and $D^{max}_s$ decreases monotonously with $\alpha$. More precisely, it is found that $D^{max}_s/t \approx 0.21$, $0.40$, $0.52$ and $0.63$ for respectively $\alpha=2$, $1$, $0.5$ and $0.1$, where $U_{max}/t \approx 5.5$, $3.1$, $2.2$ and $1.4$ respectively. In the limit of large $U$, $D_s$ is found to scale as $1/|U|$. This is expected since the physics is that of repulsive hardcore bosons whose effective mass is proportional to $|U|$ \cite{Hard-core_bosons}. Here, for $|U| \gg t$, it can be shown analytically that $D_s = \frac{2t^2}{|U|}$ (see appendix B for the details), it corresponds to the dashed line in Fig.$\,$\ref{Fig. 3}\textbf{a}. Let us finally discuss the specific case of large values of $\alpha$, e.g. $\alpha = 5$ as plotted in the inset of Fig$\,$\ref{Fig. 3}\textbf{a}. The shape of $D_s(U)$ has notably changed with respect to the cases discussed previously. The SF weight increases slowly as $|U|$ increases up to $|U|/t \approx 10$ after which it exhibits now a sudden jump before reaching its maximum at $|U|/t \simeq 14$ and beyond $|U|/t \approx 25$ it finally scales as $D_s = 2t^2/|U|$.

\subsection{Thermal fluctuation effects}

In this section, the temperature is introduced. Our main purpose is to understand how thermal fluctuations affects the superfluid weight and by extension characterize the cross-over temperature $T^*$ between the metallic (or insulating) and superconducting phases.

\begin{figure}[h!]
    \centering
    \includegraphics[scale=0.45]{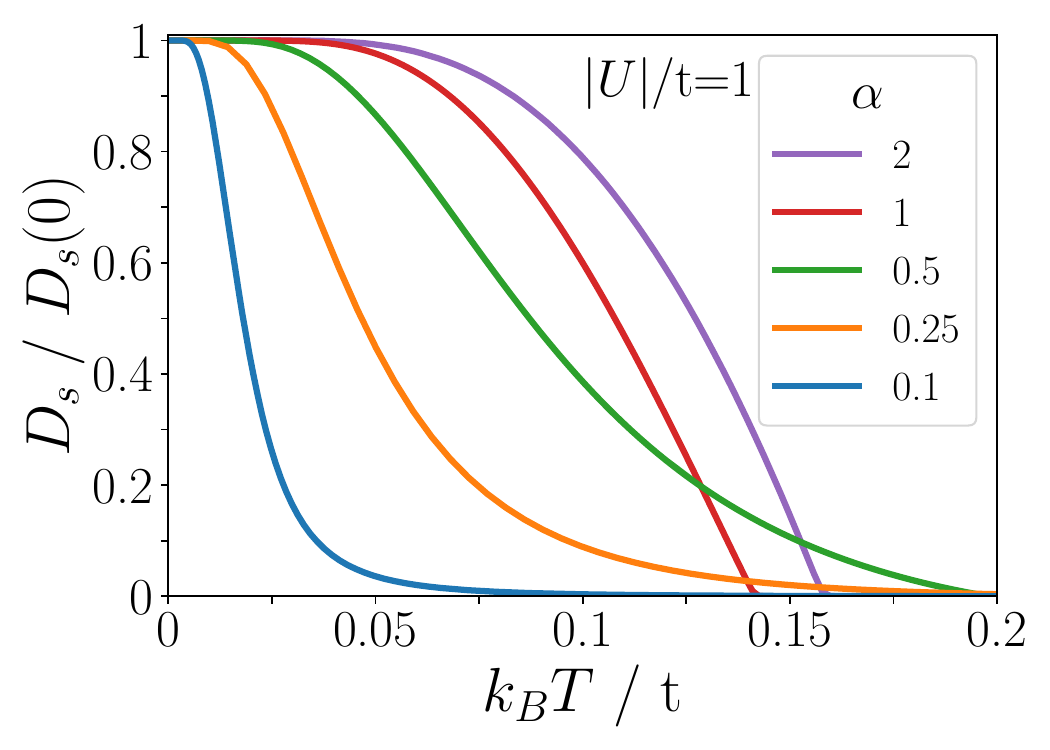} 
    \caption{Superfluid weight (rescaled by its value at $T=0$) as a function of temperature for $|U|/t = 1$ and $\alpha = 0.1, 0.25, 0.5, 1$ and $2$.}
    \label{Fig. 4}
\end{figure}
In Fig.$\,$\ref{Fig. 4} is shown $D_s$ (rescaled with respect to its value at $T=0$) as a function of $T$ for $|U|/t=1$ and different values of $\alpha$. Since $D_s(0)$ has already been discussed in Fig.$\,$\ref{Fig. 3}\textbf{a}, the discussion here focuses on the evolution of the shape and concavity of $D_s(T)$. Two different regimes are observed. First, for the largest values of $\alpha$ ($\alpha=1, 2$), $D_s(T)$ is concave and similar to a conventional BCS curve. 
As $\alpha$ decreases, an inflection point appears for $\alpha \leqslant 0.5$, and $D_s(T)$ is now convex at higher temperature. Furthermore, the region where $D_s(T) \approx D_s(0)$ is found to shrink drastically as $\alpha$ reduces. For instance, it decays by a factor $6$ when $\alpha$ varies from $2$ to $0.1$. More importantly, after the inflexion point, $D_s(T)$ exhibits a long tail before it finally vanishes. This means that the characteristic temperature for estimating the magnitude of thermal fluctuations is much lower than the BCS critical temperature.

According to Mermin-Wagner theorem, a continuous symmetry cannot be spontaneously broken at finite temperature in both one and two-dimensional systems. However, in 2D systems, a transition of topological nature can occur at finite temperature, it is known as the Berezinsky-Kosterlitz-Thouless transition \cite{Berezinsky_1972, Kosterlitz_1972, Kosterlitz_1973}. In this case, no continuous symmetry is broken and a quasi long range order below $T_{BKT}$ is established. Above $T_{BKT}$ the pair-pair correlation functions decay exponentially, and below $T_{BKT}$ they exhibit a power law decay where the exponent is T dependent. In two-dimensional superconducting systems, $T_{BKT}$ is defined as follows \cite{Formule_Tbkt, Peotta_Nature},
\begin{equation}
    D_s(T_{BKT}) = \frac{8}{\pi} k_B T_{BKT} .
    \label{Tbkt}
\end{equation}
In order to define for our one-dimensional chain a characteristic temperature $T^*$ above which the SF weight strongly reduces, we propose to use Eq.$\,$\eqref{Tbkt} as criterion. Instead, we could have chosen a different criterion such as $D_s(T^*) = 0.3 D_s(0)$ but that would only have minor effects on the following discussion. 

\begin{figure}[h!]
    \centering
    \includegraphics[scale=0.4]{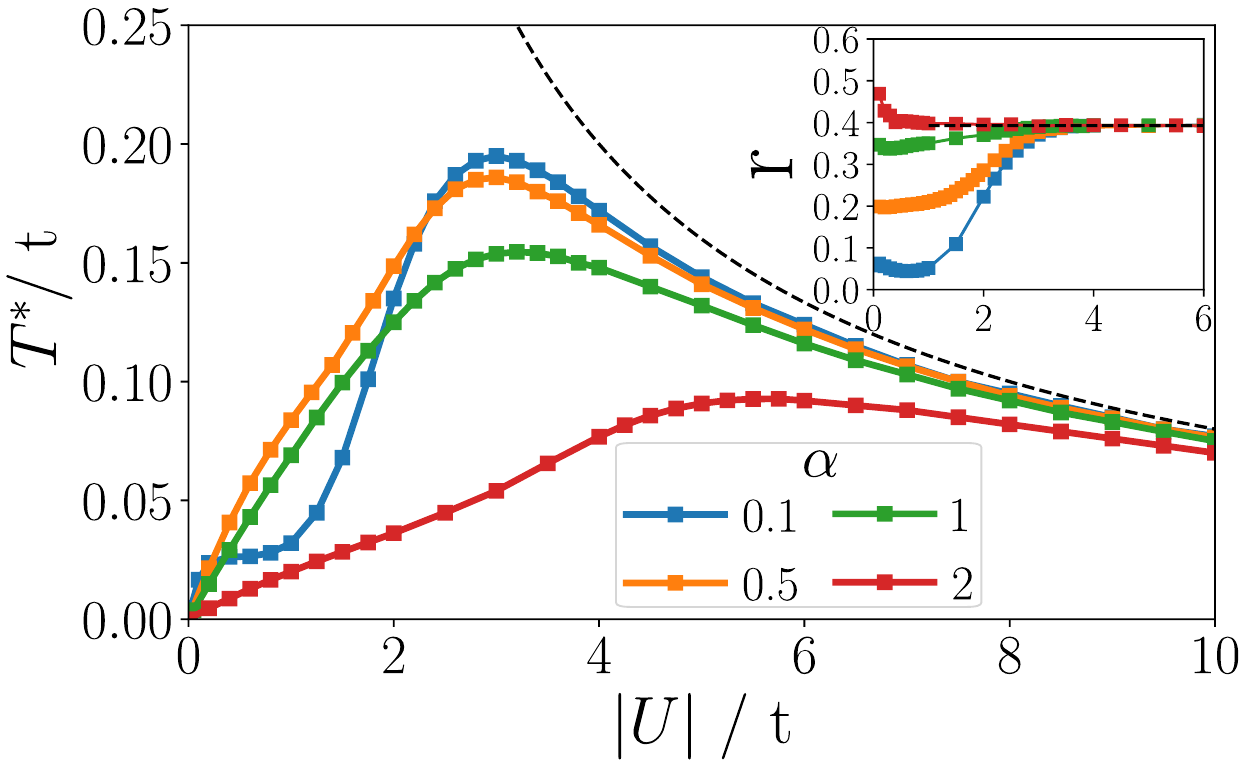} 
    \caption{Crossover temperature as a function of $|U|/t$ for $\alpha=0.1, 0.5, 1$ and $2$. The inset pictures the ratio $r = T^*/D_s(0)$.}
    \label{Fig. 5}
\end{figure}

In Fig.$\,$\ref{Fig. 5} is depicted $T^*$ as a function of the interaction strength for different values of $\alpha$. First, at weak coupling, $T^*$ scales linearly with $|U|$ \cite{Volovik_T_linear, Volovik_T_linear_Flat_bands_in_topological_media, Torma_revisiting} for any value of $\alpha$. However, as $\alpha$ increases, the slopes decrease drastically from $0.2$ for $\alpha=0.1$ down to $0.02$ for $\alpha=2$. In the mean time, the region where $T^* \propto |U|$ has tripled between $\alpha=0.1$ and $\alpha=1$, before it finally drops as $\alpha$ increases further. 
In the intermediate regime, when $\alpha \leqslant 1$, one observes an $\alpha$-independent maximum for $T^*$ located at $|U|/t \approx 3$. On the other hand, for $\alpha \geqslant 1$, it varies strongly with $\alpha$, e.g. for $\alpha=2$, the maximum location is $|U|/t \approx 5.8$.
In contrast, Fig.$\,$\ref{Fig. 3}\textbf{a} has shown that the value of $|U|$ for which $D_s(0)$ is maximum varies with $\alpha$ even when $\alpha \leqslant 1$.
Note that $T^*$ reaches its maximum after the linear region discussed above, except for the peculiar case of $\alpha=0.1$ where a clear quasi-plateau is observed for $|U|/t \in [0.2,1]$. After the plateau, $T^*$ inflates up to its maximum.
Beyond the maximum, $T^*$ decays and converge towards an $\alpha$-independent behavior. To find out how $T^*$ scales with the interaction strength, we have plotted in the inset the ratio $r = T^* / D_s(0)$ as a function of $|U|/t$. We clearly find a constant ratio $r=r_\infty=0.39$ independent of $\alpha$ when $|U|/t \geqslant 4$. However, the more $\alpha$ increases the faster $r=r_\infty$. Indeed, for $\alpha=2$, the limit is already reached for $|U|/t=1$ while for $\alpha=0.1$, $|U|/t$ must be larger than $4$. In addition, the smaller $\alpha$ is, the stronger $r$ depends on $U$ before $r=r_\infty$.
From the asymptotic scaling $D_s(0)=\frac{2t^2}{|U|}$ (Fig.$\,$\ref{Fig. 3}\textbf{a}), in the large $|U|$ limit, one finds $T^* = 0.39 \, D_s(0) \simeq 0.8 t^2/|U|$ which corresponds, in Fig.$\,$\ref{Fig. 5}, to the black dashed line.

\section{Conclusion}

To conclude, we have addressed the FB superconductivity in the one-dimensional stub chain which allows the independent tuning of the QM and of the electron-electron interaction strength $|U|$. For that purpose, within the Bogoliubov-de-Gennes approach, we  have studied in details the competition between $|U|$ and the QM $\langle g \rangle$ on the pairings and on the superfluid weight. In addition to the numerical calculations, we have provided several analytical results in both the weak and strong coupling regime. In the weak coupling regime, it is shown that the SF weight $D_s$ scales linearly with $|U|$ and exhibits two different type of behavior with respect to $\langle g \rangle$. First, when $\delta_0$, the single particle gap, is smaller than the in-chain hopping ($t$), then $D_s \propto \langle g \rangle$. On the other hand, for $\delta_0 \geqslant t$, it has been revealed that $D_s \propto \langle g\rangle ^\eta$, where $\eta \rightarrow 2$ in the limit of large gap. We have as well considered the thermal fluctuations effects on $D_s$. In particular, it is found that the shape of the SF weight depends strongly on the QM. Finally, the crossover temperature $T^*$ has been studied as a function of the interaction strength and the out-of-chain coupling. It is found that $|U|$ dependence of $T^*$ exhibits different behaviors which strongly depend on the mean value of the QM.

\section*{Appendix A: Parings in the weak coupling regime.}
\numberwithin{equation}{section}
\setcounter{equation}{0}
\renewcommand{\theequation}{A.\arabic{equation}}

Within a first order perturbation theory, the purpose of this appendix is to derive analytically 
the expression of the pairings as a function of U and $\alpha$ for the half-filled stub chain. For $q=0$, the BdG Hamiltonian reads,
\begin{equation}
\hspace{-0.2cm}
    \hspace{-0.1cm}
    \hat{H}_{BdG} = \sum_k 
    \begin{bmatrix}
        \hat{c}_{k \uparrow}^{\dagger} & \hat{c}_{-k \downarrow} \\ 
        \end{bmatrix} 
       \begin{bmatrix} 
         h_0(k)& \hat{\Delta} \\ 
         \hat{\Delta}^{\dagger}& -h_0(-k) \\
   \end{bmatrix}
          \begin{bmatrix} 
         \hat{c}_{k \uparrow} \\ 
         \hat{c}_{-k \downarrow}^{\dagger} \\
   \end{bmatrix},
%   \tag{A1}
\end{equation}
where $\hat{\Delta}=\text{diag}(\Delta_A,\Delta_B,\Delta_C)$ is the pertubation and $\hat{h}_0$ is the single particle Hamiltonian.
At half-filling, we recall that $\mu = -|U|/2$ and $n_\lambda = 1$ (uniform occupation of A, B and C sites). 

At $U=0$, the eigenstates of $\hat{H}_{BdG}$ are,
\begin{equation}
    \ket{\Psi^p_n}= 
    \begin{bmatrix}
           \ket{\phi_{n}} \\
           0 \\
    \end{bmatrix}
    \qquad
    \ket{\Psi^h_n}= 
    \begin{bmatrix}
           0 \\
           \ket{\phi_{n}} \\
         \end{bmatrix},
\label{phi-def}
\end{equation}
where $\ket{\phi_{n}}$ ($n=$ 1, 2 and 3) are the eigenvectors of $\hat{h}_0(k)$, with energy $\epsilon_n(k)$:
$\epsilon_1(k)=-\epsilon_3(k)=t d(k)$ and $\epsilon_2(k)=0$. 
In addition, the corresponding eigenvectors are $\bra{\phi_{1,3}}=\frac{1}{\sqrt{2}}(\pm1,-f_x/d,-\alpha/d)$ and $\bra{\phi_{2}}=(0,\alpha/d,-f_x/d)$, where $f_x=-2\cos(k/2)$ and $d(k)=\sqrt{f^2_x+\alpha^2}$.

Thus, the respective eigenvalues of $\ket{\Psi^p_n}$ and $\ket{\Psi^h_n}$ are $E^p_n = +\epsilon_n(k)$ and $E^h_n= -\epsilon_n(k)$.
The particle-hole symmetry of $\hat{h}_0$ implies that $E^p_n = E^h_{4-n}$ ($n = 1,2$, and $3$). For each pair of degenerate eigenstates ($\ket{\Psi^p_n}$,$\ket{\Psi^h_{4-n}}$), one can perform a first order perturbation calculation with respect to $\hat{\Delta}$. With the definition  $\delta_n=\bra{\Psi^h_{4-n}}\hat{\Delta}\ket{\Psi^p_n}$, one easily finds,
\begin{equation}
\begin{split}
\delta_1=\delta_3&= \frac{1}{2d^2} (d^2 \Delta_{A} - f_x^2 \Delta_{B}-\alpha^2 \Delta_{C}), \\
\delta_2&=-\frac{1}{d^2} (\alpha^2 \Delta_{B} + f_x^2 \Delta_{C}).
\end{split}
%\tag{A3}
\end{equation}
At the lowest order in $\Delta_\lambda$, the eigenstates of $\hat{H}_{BdG}$ are,
\begin{equation}
\ket{\Psi^\pm_n}= \frac{1}{\sqrt{2}}(\pm \ket{\Psi^p_n} + \ket{\Psi^h_{4-n}}),
\label{eqstates}
\end{equation}
with energy $E^\pm_n=\epsilon_n(k) \pm |\delta_n|$ ($n=1,2,$ and 3).

The quasi-flat band eigenstates correspond to $n=2$. At the lowest order in $\Delta_\lambda$ their dispersion is,
\begin{equation}
E^{\pm}_{2}= \pm \Bigl(\Delta_{B}\frac{\alpha^2}{d^2(k)} + \Delta_{C}\frac{f_x^2(k)}{d^2(k)} \Bigr).
\label{dispersfb}
\end{equation}
This allows the determination of the quasi-particle gap $\Delta_{QP}$. Indeed, one immediately finds, $E^\pm_2(k=0)= \pm \frac{1}{4+\alpha^2}(\alpha^2 \Delta_B+4\Delta_C)$ and $E^\pm_2 (k=\pi)= \pm \Delta_B$. As a consequence, the quasi-particle gap is located at $k=\pi$ when $\Delta_B \leqslant\Delta_C$ 
and,
\begin{equation}
\Delta_{QP}=\Delta_B.
\end{equation}
On the other hand, when $\Delta_B \geqslant \Delta_C$, the gap is located at $k=0$ and,
\begin{equation}
\Delta_{QP}=\frac{1}{4+\alpha^2}(\alpha^2 \Delta_B+4\Delta_C).
\end{equation}
In order to derive the expression of the pairings one has to recall their definition,
\begin{equation}
\Delta_{\lambda} =|U|\frac{1}{N_c}\sum_{k,n,s=\pm} \bra{\Psi^{s}_n} \hat{O}_\lambda \ket{\Psi^{s}_n} f_{FD}(E^s_n),
\label{deltas}
\end{equation}
with $\hat{O}_\lambda=\hat{c}_{k\lambda,\uparrow}\hat{c}_{-k\lambda,\downarrow}$ ($\lambda = A$,B and C) and $f_{FD}(E)=(1+e^{-\beta E})^{-1}$ is the Fermi-Dirac function. Using the expressions of $\ket{\Psi^{\pm}_n}$ as given in Eq.$\,$\eqref{eqstates}, one gets for the matrix elements the following results:
$\bra{\Psi^{\pm}_1} \hat{O}_A \ket{\Psi^{\pm}_1}=\bra{\Psi^{\pm}_3} \hat{O}_A \ket{\Psi^{\pm}_3}= \pm\frac{1}{4}$ and $\bra{\Psi^{\pm}_2} \hat{O}_A \ket{\Psi^{\pm}_2}=0$ because of the vanishing weight on A sites for the FB eigenstates.
At $T=0$, the only eigenstates which contribute to $\Delta_{\lambda}$ are $\ket{\Psi^{\pm}_3}$ and $\ket{\Psi^{-}_2}$.
Hence, at the lowest order in $|U|$ one finds, 
\begin{equation}
\Delta_{A} = 0 + O(|U|^2).
\end{equation}
Similarly one obtains,
$\bra{\Psi^{\pm}_1} \hat{O}_B \ket{\Psi^{\pm}_1}=\bra{\Psi^{\pm}_3} \hat{O}_B \ket{\Psi^{\pm}_3}= \mp \frac{1}{4} \frac{f^2_x}{f^2_x+\alpha^2}$ and $\bra{\Psi^{\pm}_2} \hat{O}_B \ket{\Psi^{\pm}_2}=\pm \frac{1}{2} \frac{\alpha^2}{f^2_x+\alpha^2}$.
The contribution from $\ket{\Psi^{+}_3}$ and $\ket{\Psi^{-}_3}$ cancel out and as expected the only non vanishing remaining contribution comes 
from the quasi-FB eigenstate $\ket{\Psi^{-}_2}$. From Eq.$\,$\eqref{deltas}, we end up with,
\begin{equation}
\Delta_{B} =\frac{|U|}{2}\frac{1}{N_c}\sum_{k} \frac{\alpha^2}{f^2_x+\alpha^2},
\label{deltab}
\end{equation}
Additionally, with the same arguments it follows,
\begin{equation}
\Delta_{C} =\frac{|U|}{2}\frac{1}{N_c}\sum_{k} \frac{f^2_x}{f^2_x+\alpha^2},
\label{deltac}
\end{equation}
For any value of $\alpha$, it implies that,
\begin{equation}
\Delta_{B}+\Delta_{C} =\frac{|U|}{2}.
\end{equation}
Finally, the sum in Eq.$\,$\eqref{deltab} can be calculated analytically leading to,
\begin{equation}
\Delta_{B} =\frac{|U|}{2}\frac{|\alpha|}{\sqrt{\alpha^2+4}}.
\label{deltab2}
\end{equation}

These expressions of $\Delta_{A}$ and $\Delta_{B}$ obtained in the limit of vanishing $|U|$ are plotted in Fig\eqref{Fig. 2}b.

\section*{Appendix B: Superfluid weight in the strong coupling regime.}
\numberwithin{equation}{section}
\setcounter{equation}{0}
\renewcommand{\theequation}{B.\arabic{equation}}

In this appendix, our goal is to calculate analytically the expression of the superfluid weight $D_s$ as a function of $U$ in the strong coupling regime. We recall that the SF weight is define as 
\begin{equation}
    D_s = \frac{1}{N_c} \frac{\partial \Omega (q)}{\partial q^2} \Big|_{q=0}, 
      \label{Ds_def} 
\end{equation}
where $\Omega(q)$ is the grand potential that reads, at T=0, 
\begin{equation}
    \Omega(q) = \sum_{k,n} E_{n}^-(k,q).
    \label{Omega_T0}
\end{equation}
In Eq.$\,$\eqref{Omega_T0}, $E_n^\pm(k,q)$ refers to the energies of the eigenstates $\ket{\Psi_n^\pm}$ of $H_{BdG}$  after the Pieirls substitution. From this expression of $\Omega(q)$, following Refs \cite{Peotta_Nature, Dice_chinois}, one find the following exact expression for the SF weight,
\begin{equation}
\begin{split}
    D_s = \frac{2}{N_c} \sum_{k,mn} &\frac{|\bra{\Psi_n^-}\frac{\partial \hat{H}_{BdG}}{\partial q}\ket{\Psi_m^+}|^2}{E_{n}^- - E_{m}^+} - \\
    & \frac{|\bra{\Psi_n^-}\frac{\partial \hat{H}_{BdG}}{\partial k}\ket{\Psi_m^+}|^2}{E_{n}^- - E_{m}^+} \Big|_{q=0},
\end{split}
\end{equation}
Let us now focus on the half-filled case. According to Lieb's theorem \cite{Th_Lieb_uniform}, the occupation is uniform $n_\lambda=1$ yielding $\hat{h}^\sigma(k)=\hat{h}_0(k)$. The only dependence on the coupling being in $\Delta_\lambda$, one can express $D_s$ as a function of the one particle velocity operator $\hat{v}_0(k) = \frac{\partial \hat{h}_0(k)}{\partial k}$ as follows,
\begin{equation}
    D_s = \frac{2}{N_c} \sum_{k,mn} \frac{|\bra{\Psi_n^-}\hat{V}\ket{\Psi_m^+}|^2 - |\bra{\Psi_n^-}\hat{\Gamma}\hat{V}\ket{\Psi_m^+}|^2}{E_{n}^- - E_{m}^+},
    \label{Ds_V}
\end{equation}
where are introduced the $6 \times 6$ matrices $\hat{\Gamma}=\text{diag}(\mathbb{\hat{I}}_{3\times 3}, -\mathbb{\hat{I}}_{3\times 3})$ and $\hat{V} = \text{diag}(\hat{v}_0,\,\hat{v}_0)$. In the strong coupling regime, all pairings are uniform i.e. $\Delta_\lambda = \Delta = \frac{|U|}{2}$ and the diagonalization of $\hat{H}_{BdG}$ gives
\begin{equation}
\begin{split}
    E_n^\pm &= \pm \sqrt{\epsilon_n^2 + |\Delta|^2}, \\
    \ket{\Psi_n^+} &= u_n \ket{\Psi_n^p} + v_n \ket{\Psi_n^h}, \\
    \ket{\Psi_n^-} &= -v_n^* \ket{\Psi_n^p} + u_n^* \ket{\Psi_n^h},
\end{split}
\end{equation}
with $|u_n|^2 = \frac{1}{2} \Big(1 + \frac{\epsilon_n}{E_n^+} \Big)$ and $|u_n|^2 + |v_n|^2 = 1$. 
$\ket{\Psi_n^p}$ and $\ket{\Psi_n^h}$ are defined in Eq.\eqref{phi-def}.
In the limit $|U| \gg t$ one finds $|u_n|^2 = |v_n|^2 = \frac{1}{2}$ and $E_n^\pm =\pm|\Delta|$. 
From this, we can give the expression of the matrix elements of Eq.\eqref{Ds_V} as a function of the one-particle Hamiltonian ($\hat{h}_0$) eigenstates $\ket{\phi_n}$,
\begin{equation}
\begin{split}
    \bra{\Psi_n^-}\hat{\Gamma}\hat{V}\ket{\Psi_m^+} &= - \bra{\phi_n}\hat{v}_0\ket{\phi_m}  \\ 
    \bra{\Psi_n^-}\hat{V}\ket{\Psi_m^+} &= 0.
\end{split}
\end{equation}
Thus, the SF weight becomes,
\begin{equation}
    D_s = \frac{2}{|U|N_c} \sum_{k,nm} |\bra{\phi_n}\hat{v}_0\ket{\phi_m}|^2.
\end{equation}
The sum coincides with $\text{Tr}[\hat{v}_0^2]$ whose value is,
\begin{equation}
    \frac{1}{N_c} \text{Tr}[\hat{v}_0^2] = t^2.
\end{equation}
This finally leads to,
\begin{equation}
    D_s = \frac{2t^2}{|U|}.
\end{equation}

\hspace{1cm}

\section*{Appendix C: Superfluid weight in the weak coupling regime}
\numberwithin{equation}{section}
\setcounter{equation}{0}
\renewcommand{\theequation}{C.\arabic{equation}}

In this appendix our purpose is to derive analytically the expression of the superfluid weight $D_s$ as a function of $\alpha$ in the limit of small $|U|$.
The calculations are done for the half-filled stub lattice and at $T=0$. Starting with the definition as given in Eq.$\,$\eqref{Ds_def} one can write,
\begin{equation}
D_s =\frac{1}{N_c} \sum_{k,n} \frac{\partial^2 E_{n}^-(k,q)}{\partial q^2}\Big|_{q=0},
\end{equation}
where $E_{n}^-(k,q)$ ($n= 1$,2 and 3) are the negative eigenvalues of the filled QP.
With the same notation as those used in the Appendix A, the eigenstates of $\hat{H}_{BdG}$ for $q \neq 0$ and $U=0$ are,
\begin{equation}
    \ket{\Psi^p_n}= 
    \begin{bmatrix}
           \ket{\phi^{q}_{n}} \\
           0 \\
    \end{bmatrix}
    \qquad
    \ket{\Psi^h_n}= 
    \begin{bmatrix}
           0 \\
           \ket{\phi^{-q}_{n}} \\
         \end{bmatrix}
\end{equation}
with respective eigenvalues $E^p_n=\epsilon_n(k+q)$ and $E^h_n=-\epsilon_n(k-q)$, where $\ket{\phi^{\pm q}_{n}}=\ket{\phi_{n}(k\pm q)}$ ($n=$ 1, 2 and 3). We recall that $\ket{\phi_{n}}$ is the eigenvector of $\hat{h}_0$, with energy $\epsilon_n$.
For a non vanishing $q$, the dispersive bands (DB) are non degenerate and $E^{p}_3 \neq E^{h}_1$, whilst the FB energy is doubly degenerate $E^{p}_2 = E^{h}_2$. 
When the perturbation $\hat\Delta$ is introduced, at first order the DB energy remains unchanged and the degeneracy of the FBs is lifted, leading to,
\begin{equation}
    \ket{\Psi^\pm_2}= \frac{1}{\sqrt{2}}
    \begin{bmatrix}
           \ket{\phi^{q}_{n}} \\
           \pm \ket{\phi^{-q}_{n}} \\
    \end{bmatrix},
\end{equation}
where the energy of these quasi-FB eigenstates is, 
\begin{equation}
E^{\pm}_{2}(k,q)= \pm \frac{1}{d_{k+q}.d_{k-q}}
\Bigl(\alpha^2 \Delta_{B} + \Delta_{C} f_x^{k+q}f_x^{k-q} \Bigr),
\label{dispersfbn}
\end{equation}
 with the notations of the Appendix A, $f_x^{k\pm q}=-2\cos(\frac{1}{2}(k\pm q))$ and $d_{k\pm q}=\sqrt{(f_x^{k\pm q})^2+\alpha^2}$.
 
 Thus, the ground-state energy per unit cell is given by,
\begin{equation}
E^{GS} (q)/N_c= \frac{1}{N_c} \sum_k (E^{p}_3 + E^{h}_1 + E^{-}_{2}).
\end{equation}

\begin{figure}[h!]
    \centering
    \includegraphics[scale=0.5]{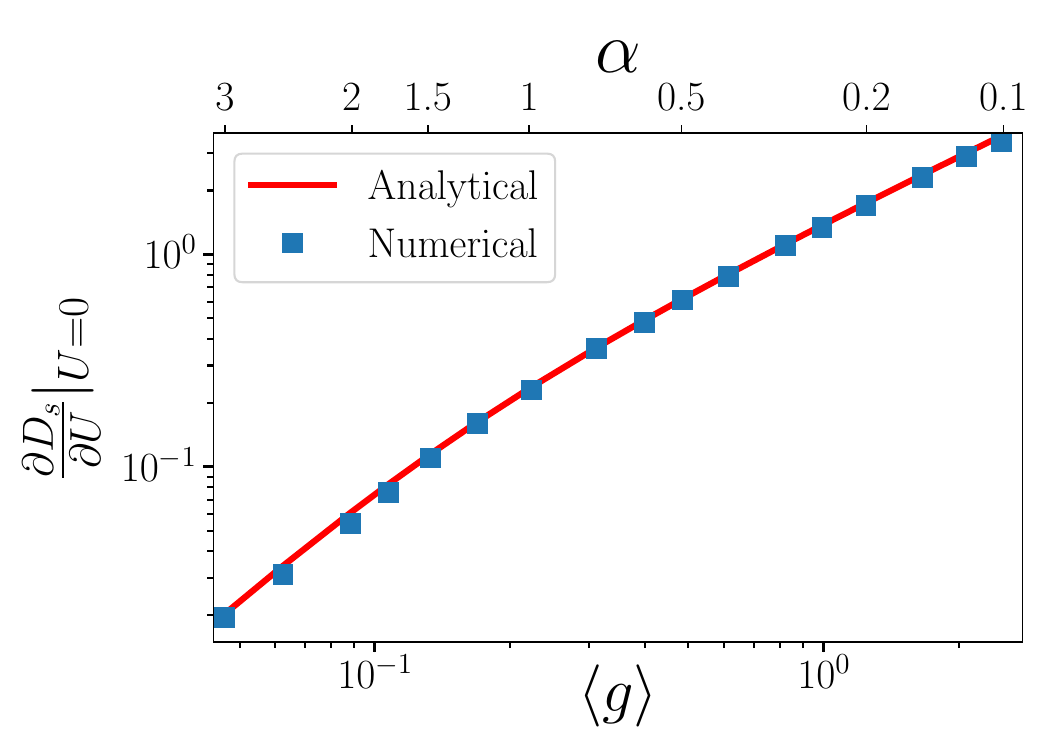} 
    \caption{$\frac{\partial D_s}{\partial U}|_{U=0}$ as a function of $\langle g \rangle$, the mean value of the quantum metric (square symbols). The corresponding values of $\alpha$ are depicted on the upper $x$-axis. The symbols are the numerical data and the continuous line is the analytical result as given in Eq.$\,$\eqref{ds-analyt}.}
    \label{Fig. 6}
\end{figure}

To get the expression of the SF weight, we now left with the calculation of the second derivative of $E^{GS}$ with respect to q.
First, notice that it can be easily shown that $\sum_k \frac{\partial^2 E^{p}_3}{\partial q^2}=\sum_k 
\frac{\partial^2 E^{h}_1}{\partial q^2}=0.$ Thus, as expected for the half-filled chain, $D_s$ depends only on the second derivative of the energy of the occupied quasi-FB. A direct double derivation with respect to q of Eq$\,$.\eqref{dispersfbn} gives,
\begin{eqnarray}
\frac{\partial^2 E^{-}_2}{\partial q^2}\Big|_{q=0} = -2\alpha^2 \Delta_B 
\Bigl(\frac{c_k}{d^4_k} + 2\frac{s^2_k}{d_k^6} \Bigr) + \nonumber \\ 
2\Delta_C \Bigl( \frac{1}{d^2_k} -2 \frac{c_k(c_k+1)}{d_k^4} 
-4 \frac{s_k^2(c_k+1)}{d^6_k} 
\Bigr),
\label{energy2}
\end{eqnarray}
where $c_k=\cos(k)$, $s_k=\sin(k)$ and $d_k=d(k)$. 
To obtain the final expression of the SF weight, one has to calculate several integrals of the form,
\begin{eqnarray}
 C_{p}^{nm}=\int_{-\pi}^{+\pi} \frac{c^n_ks^m_k} {d^p_k} \frac{dk}{2\pi}.
\end{eqnarray}
The first term in Eq.$\,$\eqref{energy2} depends on $C_{4}^{10}$ and $C_{6}^{02}$ and the second one on
$C_{2}^{00}$, $C_{4}^{10}$, $C_{4}^{20}$, $C_{6}^{12}$ and $C_{6}^{02}$. To facilitate the calculation of this set of integrals, it is convenient to define the following function, $F(u)=\frac{1}{2\pi}\int_{-\pi}^{+\pi} \frac{dk}{u + \cos(k)}$. It can be shown (standard residue calculation) that for $u > 1$, $F(u)=\frac{1}{\sqrt{u^2-1}}$. Using F and its derivative F', and after some lengthy but straightforward steps, one finds,
\begin{eqnarray}
C_{2}^{00}&=&\frac{1}{2}F(\eta), \\
C_{4}^{10}&=&\frac{1}{4}(F(\eta)+\eta F'(\eta)), \\
C_{4}^{20}&=&\frac{1}{4}(-\eta F(\eta) - F'(\eta)+1), \\
C_{6}^{02}&=&-\frac{1}{16}(F(\eta) +\eta F'(\eta)), \\
C_{6}^{12}&=& \frac{1}{8}(\frac{1}{2}F'(\eta) +\eta F(\eta)-1),
\end{eqnarray}
where for practical reason the variable $\eta = 1 +\frac{\alpha^2}{2}$ is introduced. 

After inserting in Eq.$\,$\eqref{energy2} the $C_{p}^{nm}$'s given above, we finally end up with the analytical expression of the SF weight,
\begin{eqnarray}
D_s =-\frac{1}{4}\alpha^2 \Delta_B 
\Bigl(F(\eta) +\eta F'(\eta) \Bigr) + \nonumber \\ 
\frac{1}{2}\Delta_C \Bigl( F(\eta) + F'(\eta) -\eta F'(\eta)
\Bigr).
\label{ds-analyt}
\end{eqnarray}
Using the expressions of $\Delta_B$ and $\Delta_C$ as given in the Appendix A, we have been able to compare this analytical expression of $D_s$ with the numerical data. The result depicted in Fig.$\,$\ref{Fig. 6} reveals an excellent agreement between the numerical and analytical data for the whole range of values of $\alpha$.

From Eq.$\,$\eqref{ds-analyt} one can now extract the asymptotic behaviour of the SF weight for both limits: (1) $\alpha \ll 1$ which correspond to large values of the QM and small one-particle gap and (2) $\alpha \gg 1$ for which the QM is small and the gap is large.

In the first case, one gets,
\begin{eqnarray}
D_s = \frac{3}{8\alpha} |U| = \frac{3}{2} |U| \langle g \rangle
\end{eqnarray}
On the other hand, in the second one ($\alpha \gg 1$) one finds,
\begin{eqnarray}
D_s = \frac{3}{\alpha^4} |U| = 12 |U| \langle g \rangle^2
\end{eqnarray}
The SF weight scales linearly with $\langle g \rangle$ when $\alpha \ll 1$ and as $\langle g \rangle^2$ when $\alpha \gg 1$.

\end{document}